\documentclass[fleqn,usenatbib]{mnras}

\usepackage{newtxtext,newtxmath}

\usepackage[T1]{fontenc}

\DeclareRobustCommand{\VAN}[3]{#2}
\let\VANthebibliography\thebibliography
\def\thebibliography{\DeclareRobustCommand{\VAN}[3]{##3}\VANthebibliography}

\usepackage{graphicx}	
\usepackage{amsmath}	
\usepackage[T2A]{fontenc}
\usepackage[utf8]{inputenc}
\usepackage[english]{babel}
\usepackage[normalem]{ulem}  
\usepackage{xcolor}

\newcommand{\noopsort}[1]{}

\title[Stabilization of astropause]{Stabilization of the astropause by periodic fluctuations of the stellar wind}

\author[S.D. Korolkov]{
S. D. Korolkov,${}^1{}^2{}^3$\thanks{E-mail: korolkov.msu@mail.ru}
V. V. Izmodenov,${}^1{}^2{}^3$\thanks{E-mail: izmod@iki.rssi.ru}
\\
$^{1}$Space Research Institute (IKI) of Russian Academy of Sciences, Moscow, Russia\\
$^{2}$Lomonosov Moscow State University, Moscow center for fundamental and applied mathematics, Moscow, Russia\\
$^{3}$HSE University, 20 Myasnitskaya Ulitsa, Moscow 101000, Russia
}

\date{Accepted XXX. Received YYY; in original form ZZZ}

\pubyear{2022}

\begin{document}
\label{firstpage}
\pagerange{\pageref{firstpage}--\pageref{lastpage}}

\maketitle

\begin{abstract}

The main goal of the paper is to explore why observations of many astrospheres (or circumstellar bubbles) show quite stable and smooth structures of astropauses - the tangential discontinuities separating the stellar and interstellar winds, - while both theory and numerical simulations suggest that tangential discontinuities are unstable due to well known Kelvin-Helmholtz (K-H) instability. It was recognized before that magnetic fields may stabilize the astropauses. In this paper, we explore another mechanism to reduce the K-H instability of the astropauses. This mechanism is a periodic change of the stellar wind dynamic pressure. Fluctuations of the stellar wind parameters are quite expected. For example, the Sun has an 11-year cycle of global activity although there are also shorter periods of the solar wind fluctuations.
   
We performed the parametric numerical study and demonstrate that the development of the K-H instability depends on the dimensionless parameter $\chi$ which is the ratio of the stellar wind terminal speed and interstellar flow speed. The larger the parameter $\chi$,  the larger the fluctuations caused by the K-H instability. It has been shown that the K-H instability is convective which agrees with the previous linear analysis.

The stabilization of the astropause by the periodic fluctuations in the stellar wind lead is demonstrated.
It is shown that for the solar wind the most effective stabilization occurs when the period of stellar parameter change is about 1-4 years. For the eleven year solar cycle, the stabilization effect is weaker.
\end{abstract}

\begin{keywords}
Solar wind -- Instabilities --
                Sun: evolution, heliosphere --
                Methods: numerical
\end{keywords}



\section{Introduction}

\begin{figure}
	\begin{minipage}[h!]{1.00\linewidth}
		\center{\includegraphics[width=0.9\linewidth]{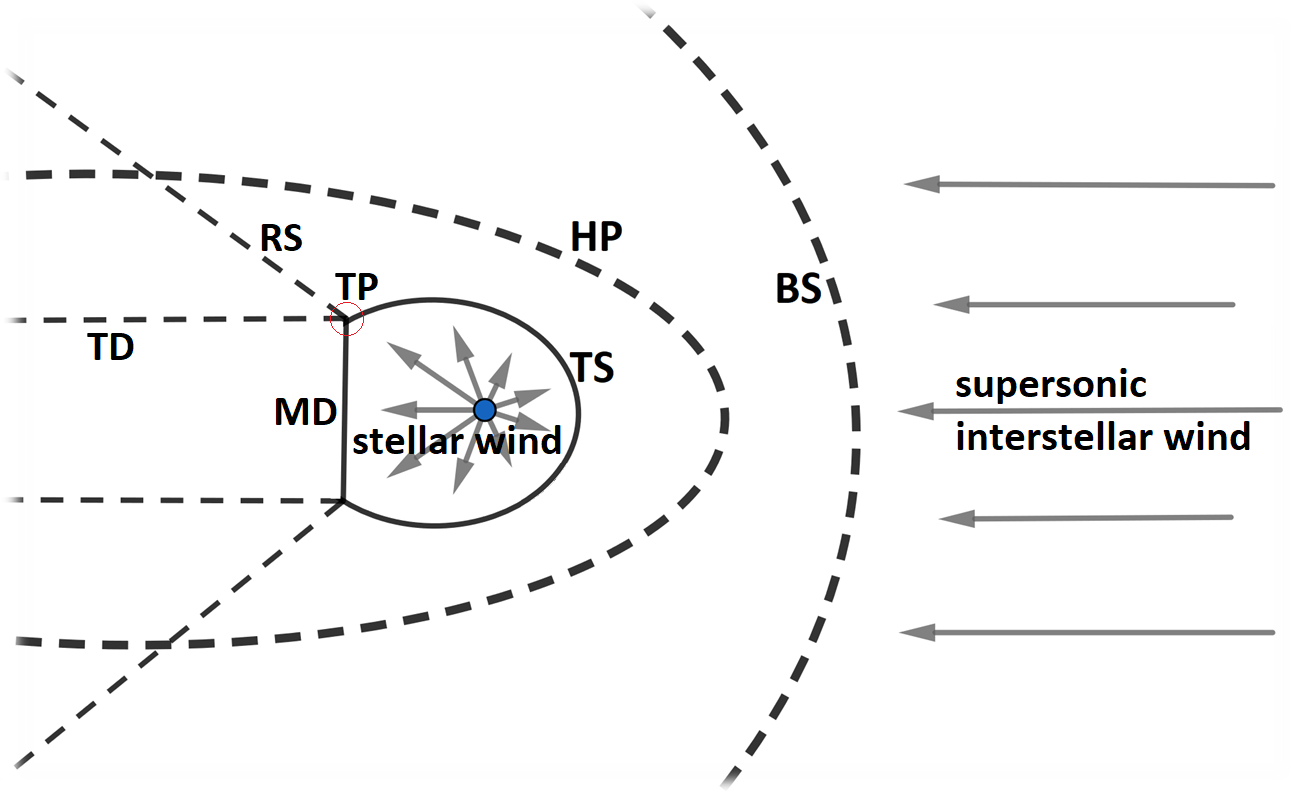} }
	\end{minipage}
	\caption{Qualitative picture of the supersonic stellar wind interaction with the supersonic flow of the interstellar medium. TS - termination shock, HP - heliopause, BS - outer bow shock, MD - Mach disk, RS - reflected shock wave, TD - secondary tangential discontinuity, TP -  triple point.}
	\label{page_1}
\end{figure}

We consider the problem of hypersonic stellar wind interaction with the interstellar medium. A schematic picture of this interaction is shown in Fig.~\ref{page_1}. A tangential discontinuity (HP), called astropause (or heliopause in the case of the Sun), separates the stellar wind from the interstellar medium. Existence of the heliopause has been proved when Voyagers 1 and 2 crossed it at distances of 122 and 118 AU in 2012 and 2018, respectively. For other stars, the astropause-like structures were observed in the infrared or H-alpha images (see, e.q., \cite{Kobul_2016, Kobul_2017}).

To be able to slow down and flow around the astropause, the supersonic flow of the stellar wind must pass through a shock that is called the termination shock (TS). For the heliosphere, the TS has been crossed by Voyagers at 94 and 85 AU in 2004 and 2007, respectively.

If the relative velocity of the star and circumstellar interstellar medium is sufficiently high,  the formation of an external shock wave (bow shock - BS) can be expected. Such structures are actually observed (see, e.q., \cite{Buren1988, Buren1995}).  Further, we will pay attention to rapidly moving stars, around which the bow shock is formed. 


The problem of the stellar wind interaction with the supersonic interstellar wind has been considered firstly by \cite{baranov70} in the thin layer approximation. A qualitative picture of the flow with two shocks and tangential discontinuity between them was proposed in this paper (see Fig.~\ref{page_1}). Later, in the same approximation, a simple analytical solution was also obtained \citep{Wilkin}. The flow pattern in the tail region of interaction in the gas-dynamic case (without the influence of neutral atoms) becomes similar to the problem of the interaction of the stellar wind with the cometary ionosphere \citep{Wallis1976}. There are a Mach disk (MD), a tangential discontinuity (TD), and a weak reflected shock (RS) originated at the triple point (TP). Modern models of the astrosphere are more complex due to the need to substantiate new observational data (see, e.q., review of \cite{Herbst}).

Since the tangential discontinuity or astropause is one of the key elements that determine the flow pattern in the astrosphere then the question of its stability becomes important.  It is known from the textbooks that the tangential discontinuities are unstable due to the Kelvin-Helmholtz (K-H) instability (\cite{Helmholtz}; \cite{Kelvin}). This instability develops due to the difference in the tangential velocities at the discontinuity. The linear analysis shows that the shorter the wavelength of disturbance, the faster the growth of the wave amplitude.


The stability of heliopause/astropause has been studied in many papers. We review only some of them. \cite{Baranov1992} in a first-order approximation of the WKB method investigated the stability of the heliopause in the absence of magnetic fields for the flanks and the head region (not very close to the nose). It is shown that the head region of the heliopause is unstable to all surface waves, and the flanks, where the flow is supersonic, are unstable to waves with a tilt angle greater than the critical one. \cite{Chalov1996} also showed, using the WKB method, that the nose part of the heliopause is unstable to short-wave disturbances.

\cite{Dgani} conducted a linear stability analysis (WKB approximation) of an astropause in the assumption of isothermal flow. It has been shown that astropauses are generally unstable, and the ratio of the wind terminal velocity, $V_0$, to the speed of the interstellar medium $V_\infty$ (or star speed) characterizes the properties of instability. Interestingly, the instability manifests itself more strongly for a slow stellar wind (at $V_0 << V_\infty$).

Numerical modeling results also confirm that the heliopause/astropause is unstable (see, e.g., \cite{Wang1998}, \cite{Korolkov_2020}). 
Nevertheless, the existence of the numerical solutions means that the instability is convective, not absolute.  Convective instability means that at any given point the perturbations do not grow with time, and they increase while a fluid parcel moves along the tangential discontinuity from the head to the flanks \citep{Kulikovski1977}. \cite{Ruderman} investigated the instability of a tangential discontinuity in incompressible fluids using the Fourier-Laplace transform and have shown that the near flanks of the heliopause, where the flows on both sides are strongly subsonic, are absolutely stable, but convectively unstable.

The observations of some astrospheres confirm theoretical predictions on the convectively unstable astrospheres. One such example is the astrosphere around R Leo  \citep{Cox_2012}. Other observations demonstrate surprisingly stable astrospheric structures. For example, the K-H instability is not observed around the astrospheres of  Betelgeuse \citep{Noriega} and IRC-10414 \citep{Meyer}. Therefore,  there are some physical mechanisms of flow stabilization that are not taken into account in the gas dynamical models mentioned above. For example, \cite{Ruderman1993} found that magnetic fields can have a strong stabilizing effect, while the heliopause must still remain unstable in the flanks.

 \cite{Decin}  studied the astrosphere around Betelgeuse and noted the absence of large-scale unstable structures that are observed in numerical hydrodynamic calculations. The stabilization effect was explained by the possible influence of the interstellar magnetic field and/or the presence of an inhomogeneous mass loss process. The mechanism of mass loss is represented by alternating turning on and off (decreasing by a factor of 1000) the stellar wind. One can question whether such large variations in the mass-loss are possible.
 
\cite{Meyer2021} also noted the possibility of flow stabilization in the presence of an external magnetic field if the local interstellar medium (LISM) flow and magnetization directions differ by more than a few degrees (> 5).

It should be noted that other mechanisms of flow stabilization are possible. For example, \cite{Meyer2014} has shown that the ionization of stellar winds by external sources of radiation can result in its acceleration by a factor of 2. This effect leads to a reduction of the shear in velocities on both sides of astropause and, therefore, to a more stable astrosphere.
Such an effect applies to the astrospheres of red supergiants.

In this paper, we extend the work of \cite{Decin} and explore the influence of variations of the stellar wind parameters on the instability of the astropause. We assume that the stellar mass flux varies periodically. 
Periodic changes in the stellar parameters are observed. For example, the 11-year solar activity cycle is very well known. Analysis of solar-wind data collected by spacecraft shows other periods,  for example, a 1.3-year periodicity (see \cite{Richardson1994}).

The idea of stabilizing a system by introducing forced oscillations has already taken place in classical mechanics. This method is best known for the Kapitza's pendulum problem \citep{Kapitza1, Kapitza2, Landau1}. The unique feature of the Kapitza's pendulum is that the vibrating suspension can cause it to balance stably in an inverted position, with the bob above the suspension point. 

The structure of the paper is following. In Sect.~\ref{M} the model is formulated. Dimensionless parameters are discussed as well as the methods employed to obtain the numerical solutions.
Sect.~\ref{R} presents and discusses the results of numerical calculations. Sect.~\ref{C} summarizes the work and discusses possible future research.

\section{Model}
\label{M}

In our model, both the stellar wind and the interstellar medium are considered ideal non-heat-conducting gases with constant heat capacity and $\gamma = 5/3 $. For simplicity, the influence of magnetic fields is neglected and a pure gas dynamic approximation is considered, so Euler equations are solved. 
 
The coordinate system is centered in the Sun. The $x$ axis is directed against the incoming flow. The boundary conditions at $x \rightarrow \infty$ are the following: the incoming interstellar flow is considered parallel with density $\rho_\infty$, pressure $p_\infty$, and velocity  ${V}_\infty = - V_\infty \bf{e_x}$.

The stellar wind is assumed a hypersonic (Mach number M$>>$1) spherically symmetric source of gas. The hypersonic stellar wind is determined by two parameters: the stellar mass loss rate, $\dot{M}_\odot$, and the terminal velocity, $V_0$. We assume periodic fluctuations of the mass loss rate: $\dot{M}_\odot = \dot{M}^\mathrm{st}_\odot \cdot \left(1 + A\cdot \mathrm{cos} (2\pi * t/T_0)\right) $, where $T_0$ is the period of fluctuations and the parameter $A$ determines the magnitude of fluctuations.

The solution for the stationary boundary conditions (A=0) depends on six dimensional parameters. Two additional parameters appear for periodic boundary conditions. It is convenient to formulate the problem in dimensionless form. Let us choose $R_* =\left (\frac{1}{4 \pi} \frac{V_0 \dot{M}^{\mathrm{st}}_\odot}{\rho_\infty V_\infty^2}\right)^{1/2}$ as a characteristic distance, $t_* = R_*/V_\infty$ as a characteristic time,
$\rho_\infty$ as a characteristic density.
To give an example, for the case of the Sun  $V_0$= 450 km/s, $ \dot{M}^{\mathrm{st}}_\odot = 1.5 \cdot 10^{12}$ g/s, $V_ {\infty}$ =26.4 km/s, $n_\infty$= 0.1 cm$^{-3}$, and, therefore, $R_* = 144$ a.e., $t_* = 36$ years. For the solar cycle fluctuations $T_0 = 11$ years. 
It is easy to show (see \cite{Korolkov_2020}) that the solution to the problem is determined by five dimensionless parameters: (1) the ratio of the speeds of the stellar and interstellar winds $\chi = V_0/V_\infty$, (2) the Mach number ($M_\infty$) in the interstellar medium, (3) the parameter $\gamma$, which is 5/3 for a fully ionized plasma (it is considered constant and does not vary), (4) the period of density fluctuations in the stellar wind $ \hat{T}_0 = T_0/t_* $ (the hat is omitted below), and (5) amplitude of fluctuations $A$.



The geometric pattern of a stationary solution does not depend on the parameter $\chi $ (\cite{Korolkov_2020}). The parametric study of the stationary solutions of the problem depending on the interstellar Mach number are presented in \cite{Korolkov_2020}. 



All calculations are performed on Cartesian grids with high resolution, which is achieved by using CUDA technology of parallel programming on video cards. In numerical solution, we employ the approximate HLLC method (\cite{Gurski}, \cite{Miyoshi}). The minmod limiter (TVD scheme) has been used to improve the accuracy of the scheme. The resolutions of the computational domain were chosen: $3584 \times 2200$ and $1792\times1536$ (this is specified in the signature for each calculation), however, other resolutions were also checked. The dependence of the numerical solution on the method and grid resolution was studied in the work \cite{Korolkov_2020}.

\section{Results}
\label{R}
\subsection{Influence of the parameter $\chi$ on stability of the astropauses}


\begin{figure*}
	\begin{minipage}[h!]{0.24\linewidth}
		\center{\includegraphics[width=1\linewidth]{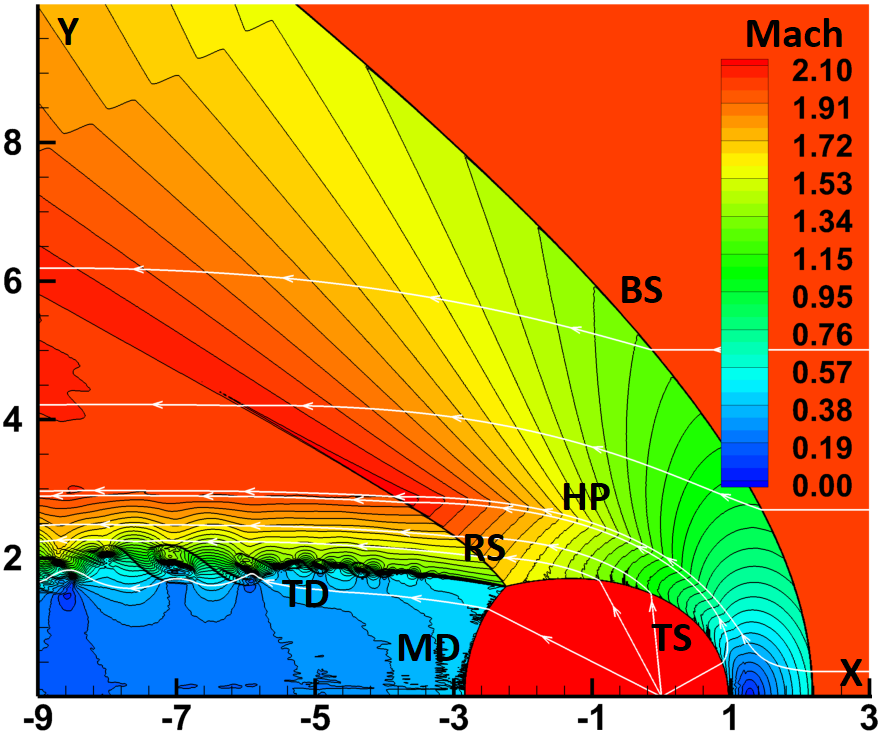}} (a) $\chi = 1.6$
	\end{minipage}
	\hfill
	\begin{minipage}[h!]{0.24\linewidth}
		\center{\includegraphics[width=1\linewidth]{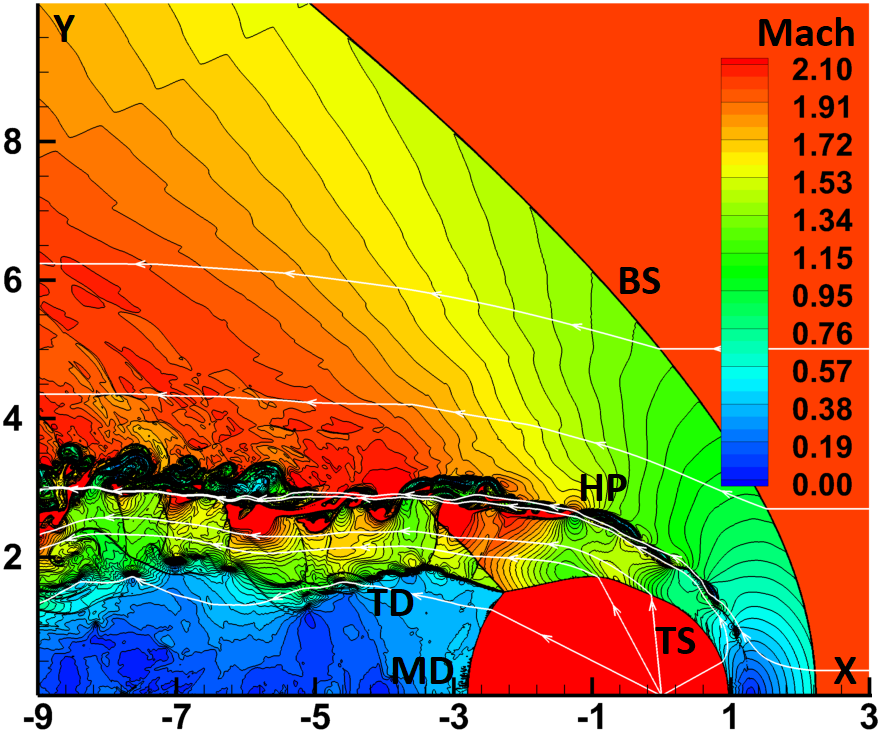}} (b) $\chi = 5$
	\end{minipage}
	\hfill
	\begin{minipage}[h!]{0.24\linewidth}
		\center{\includegraphics[width=1\linewidth]{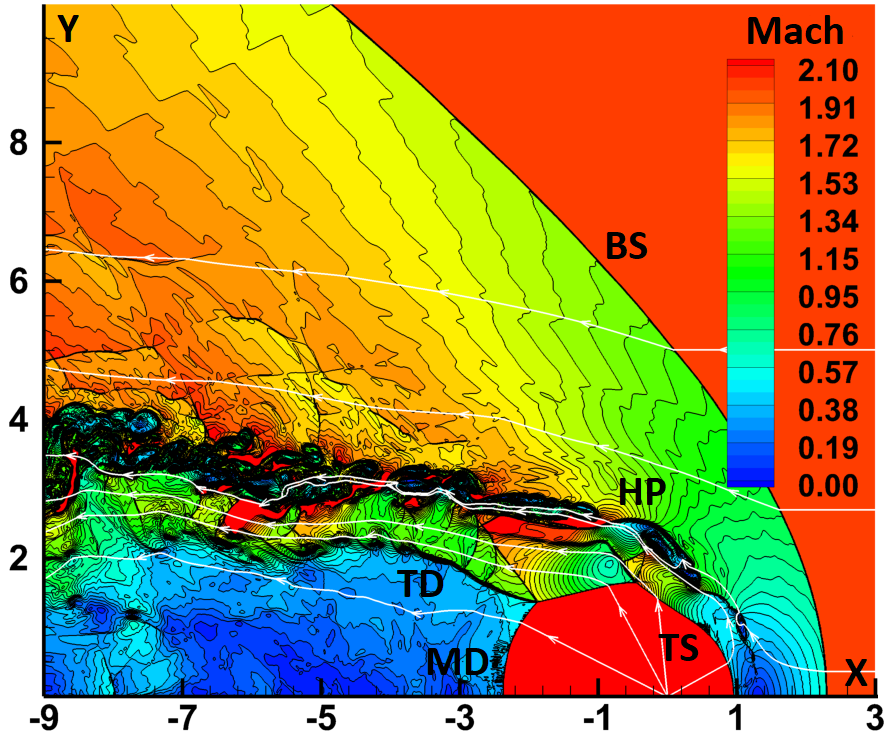}} (c) $\chi = 10$
	\end{minipage}
	\hfill
	\begin{minipage}[h!]{0.24\linewidth}
		\center{\includegraphics[width=1\linewidth]{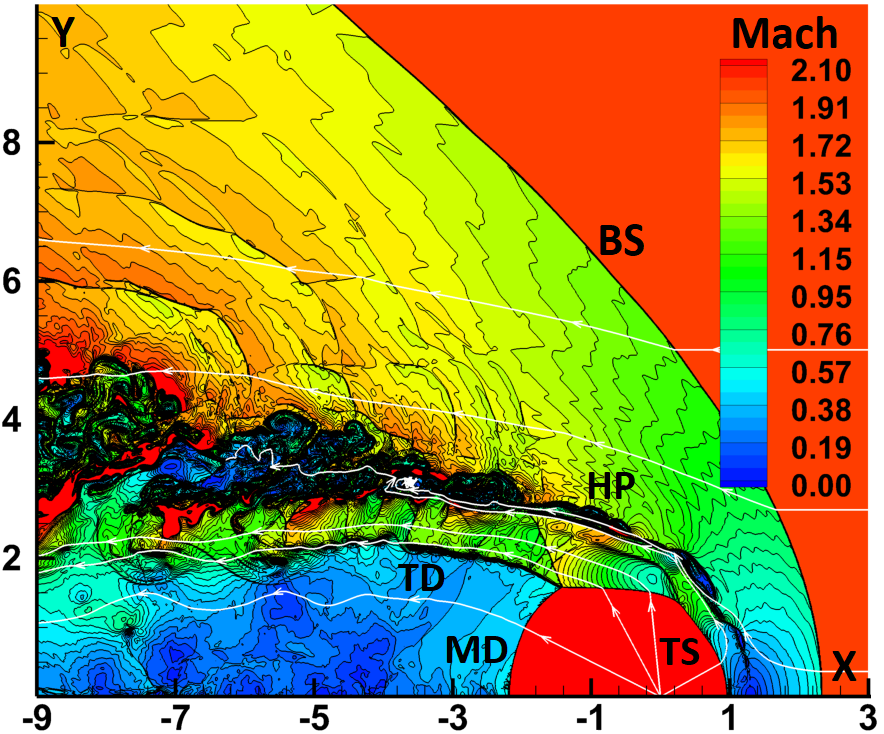}} (d) $\chi = 17$
	\end{minipage}
	\caption{Isolines of the Mach number and few streamlines for different values of the parameter $\chi$. $M_\infty = 2 $, HLLC + TVD method. The grid resolution: $2816\times2048$.}
	\label{2}
\end{figure*}

\begin{figure*}
	\begin{minipage}[h!]{1.00\linewidth}
		\center{\includegraphics[width=1.0\linewidth]{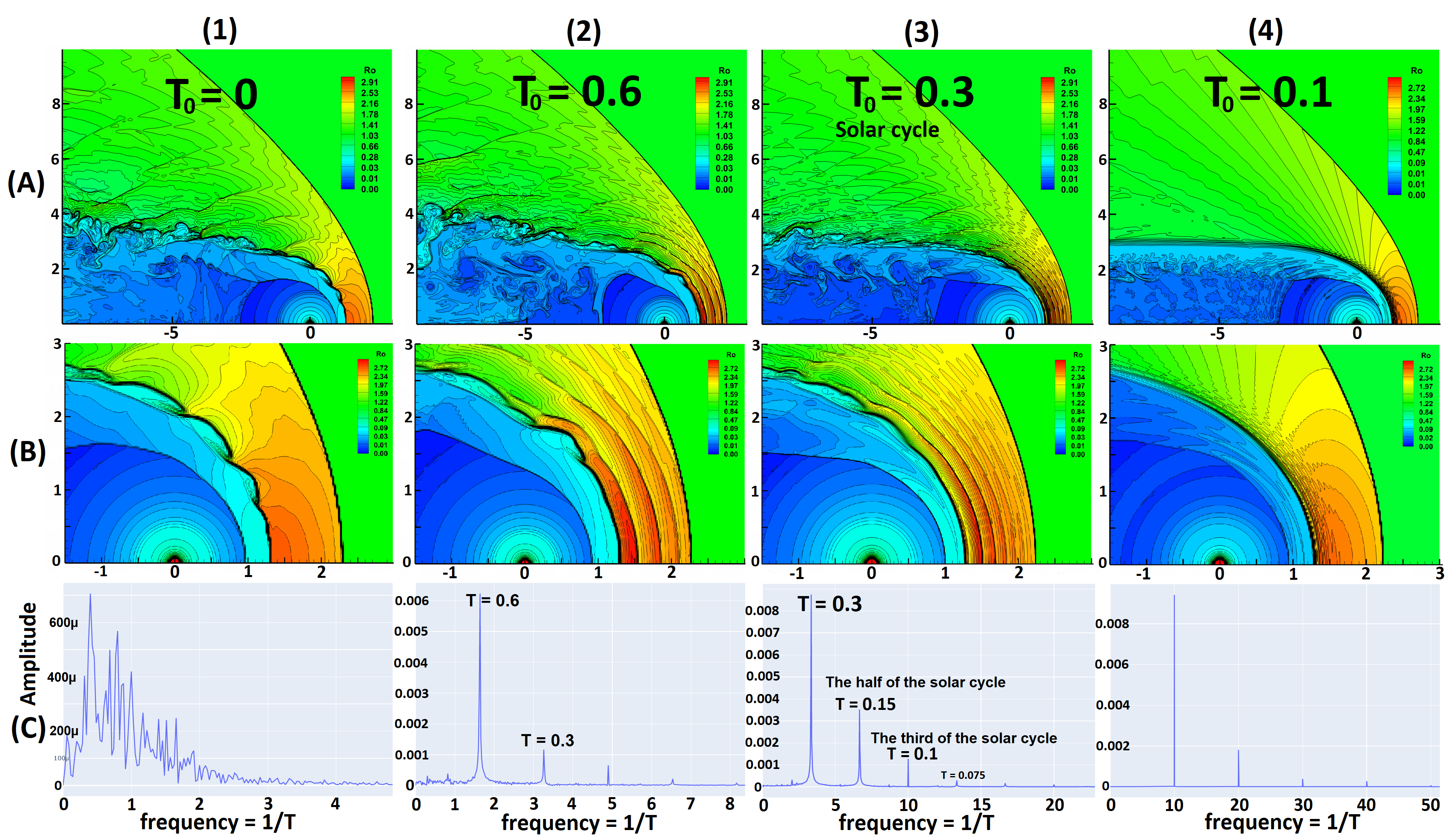} }
	\end{minipage}
	\caption{Isolines of the density (A), (B) (B is zoom A) for different values of the stellar wind period. The Fourier series expansion of the density at a point (0.7; 1.1) with time (C). The x-axis represents the oscillation frequency, and the y-axis represents the modulus of the complex amplitude. $M_\infty = 2 $, $\chi = 10$, HLLC + TVD method. The grid resolution: $1792\times1536$. The symbol <<$\upmu$>> denotes $10^{-6}$.}
	\label{3}
\end{figure*}

Firstly, we consider the problem with the stationary boundary conditions. As noted above, in this case, the problem depends on two dimensionless parameters: (1) $M_\infty$, and (2) $\chi$. In this subsection, we present results of numerical calculations with four different values of $\chi$ (a-d) and with fixed $M_\infty = 2$ (see Fig~\ref{2}).
It should be emphasized that we solve non-stationary Euler equations and it is not necessary that the obtained solutions are stationary. 
 Fig.~\ref{2} shows snapshots of the isolines of the Mach number at arbitrary chosen time-moments. Few streamlines are also shown as white curves. We have also attached a video to the calculations (see Supplement materials). The video shows fluctuations in the Mach number and density over time, as well as changes in the Mach number over time at two points in the inner and outer shock layers.
 
Panel (a) shows the results of calculations with the parameter $\chi = 1.6$. The classic flow pattern described in fig.\ref{page_1} is seen here. In the head region, the obtained solution does not vary with time. It is stationary and stable. Some perturbations of the secondary tangential discontinuity are seen in the tail region. We conclude that for the values of $\chi \sim 1$ the astropause does not fluctuate much, i.e. the instability does not develop.


For $\chi = 5$  fluctuations of the astropause appear (see panel (b) in  Fig.~\ref{2} and video "(b)-1.avi", "(b)-2.avi" in Supplement materials). It is clearly seen that the fluctuations appeared at some distance from the stagnation point in the head region.  
Also, the formation of an additional point of break of the TS becomes noticeable. The TS is not smooth (having a break) at this point.
The shock wave is not stationary. In particular, the point of break moves across the TS. We can call this change quasi-periodic with the main period of about 1.3-1.5 dimensionless time (about 46-54 years for the heliosphere). Fluctuations of the Mach number at two arbitrary points of the inner and outer shock layers reach 16 and 7\%, respectively (see "(b)-2.avi" in Supplement materials).

Panels (c, d) in the Fig.~\ref{2} show the results for $\chi = 10 $ and $\chi = 17$, respectively. $\chi=17$ is close to the value for the heliosphere.  The non-stationary fluctuating behavior of astropause is more pronounced. The fluctuations become essentially larger. The break in the TS shape at RP becomes stronger. A complex of secondary reflected shocks appears in the tail. Fluctuations of the Mach number at two arbitrary points of the inner and outer shock layers reach 29 and 10\%, respectively, for the case of $\chi = 10$ (see "(c)-2.avi" in Supplement materials), and 34 and 20\%, respectively, for the case of $\chi = 17$ (see "(d)-2.avi" in Supplement materials).

The physical reason for the observed fluctuations in the obtained solutions is clear. This is Kelvin-Helmholtz instability that appeared at the tangential discontinuities. In the standard scheme of the flow pattern shown in Fig.~\ref{page_1} there are two tangential discontinuities: (1) the astropause, and (2) the tangential discontinuity emanating from the triple point in the tail region. Obtained numerical solutions demonstrate instabilities of both.

A simple linear analysis of the stability of a tangential discontinuity between two incompressible fluids gives \citep{Landau6}:

$$ \omega = \triangle U \cdot k \cdot \dfrac{\rho_1 \pm i \sqrt{\rho_1 \rho_2} }{\rho_1 + \rho_2},$$

where $\triangle U$ - is the jump of velocities at the discontinuity, $k$ - is the wave number, $\rho_1,\ \rho_2$ - are densities of the medium on both sides of the discontinuity, $\omega$ - is the oscillation frequency.

As it is seen from this simple formula, for all possible values of parameters there are $\omega$ with a positive imaginary part. Such $\omega$ determines the exponential growth of disturbances with time. The growth of the perturbation wave is faster for the larger values of the $\triangle U$. Our numerical modeling presents the results in the non-linear problem. However, the results qualitatively are the same. The larger is $\chi$, the larger the jump of velocities at the discontinuity, and the larger the level of observed fluctuations. 
It is now absolutely clear that the parameter $\chi$ directly affects how fast the K-H instability develops.  It is important to note that the instability, in this case, is convective, so it does not destroy the solution. The fluctuations are growing, but at the same time are moving convectively out from the computational domain.  The convective behavior of heliopause has been established by \cite{Ruderman}.

\subsection{The effect of flow stabilization by a periodic stellar wind}

\begin{figure*}
	\begin{minipage}[h!]{0.8\linewidth}
		\center{\includegraphics[width=1.0\linewidth]{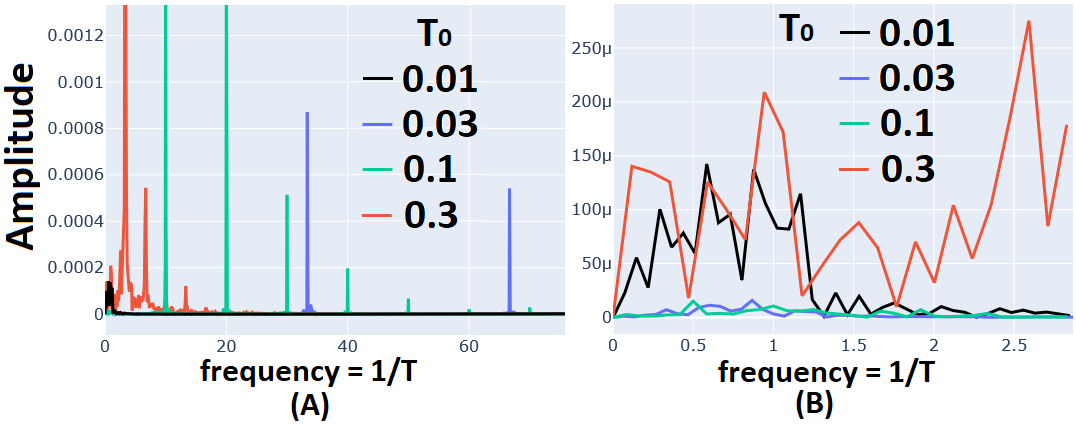} }
	\end{minipage}
	\caption{The Fourier series expansion of function of the density versus time at the point (0.7; 1.11). The x-axis represents the oscillation frequency, and the y-axis represents the modulus of the complex amplitude. $M_\infty = 2 $, $\chi = 17$, HLLC + TVD method. The grid resolution: $1792\times1536$.}
	\label{4}
\end{figure*}


In this section, we present the results of modeling with the periodic inner boundary conditions. The results are presented in Fig.~\ref{3} which has three rows and four columns. Each column corresponds to a certain value of the solar-wind period $T_0$. $T_0$ = 0 corresponds to the solutions with stationary boundary conditions. Panels in the top row (A) show the isolines of density. Middle row (B) shows the same results zoomed for the head region. The calculations performed for $\chi = 10$ and $M=2$. Bottom-row panels will be described later.

The calculations show that the periodic variations of $\dot{M}_\odot$ lead to the periodic motion of the TS and BS.  TS moves in and out with an amplitude of 0.04-0.07 in dimensionless units (7-10 AU for the heliosphere). The amplitude of the HP fluctuation is smaller. It is 0.01-0.02 in dimensionless units (2-3 AU for solar system). The BS does not move at noticeable scales. Also, since the HP acts as a piston moving in and out in the interstellar gas, a series of shock waves appear in the region between the HP and BS.
Similar results have been reported before (e.g, in the case of the heliosphere see \cite{Izmodenov2005}).

However, the most interesting is the effect of the so-called stabilization of the flow. The effect is clearly seen by comparing panels in columns (1) and (4) of Fig.~\ref{3}. The flow pattern shown in column ($T_0=0$ (1) is wavy and unstable, while the waves and instability disappear in the flow with $T_0 =0.1$. Panels (2) and (3) show intermediate cases when fluctuations are still seen but they are somewhat less pronounced than for $T_0 = 0$. 

To better characterize the fluctuations obtained in different solutions
we performed the decomposition of density as a function of time at an arbitrarily chosen point (with coordinates  (0.7; 1.11)) into a Fourier series (see row C in Fig.~\ref{3}).
The chosen point is located in the inner shock layer between the TS and HP. The x-axis shows the frequency (1/period) of each harmonic, the y-axis shows its amplitude (density in dimensionless variables). Note that we used the Fourier series in exponential form and showed the modulus of the complex amplitude.

Panel (C1) demonstrates the result of decomposition in the case of  $T_0 = 0$.
Many harmonics are observed in the flow. The largest amplitudes appear at low frequencies. For the panel (C2) with $T_0 = 0.6$, the largest amplitudes corresponds to the oscillation periods of $ T_0 $, $ T_0 / 2 $, $ T_0 / 3 $, etc. However, “turbulent” waves with other frequencies, which disturbs the flow, still exist, and their amplitudes are noticeable. The flow of $T_0 = 0.3$ (C3) is much more stable and there are almost no "turbulent" frequencies. In the last picture (C4 where $T_0 = 0.1$), only frequencies that are multiples of the wind period remain. The "turbulent" waves with other frequencies disappear.
 The effect of flow stabilization due to periodic changes in the boundary condition is clearly seen.


Now we consider the flow for $\chi = 17$. The flow with $T_0 = 0$ is more turbulent than one for $\chi = 10$. We performed calculations with $T_0$ = 0.3, 0.1, 0.03, 0.01 and performed decomposition into a Fourier series as it was described above. The results of decomposition are presented in Fig.~\ref{4}. Panels A and B have different scales (high (A) and low (B) frequency harmonics). We notice low-frequency noise (B) for cases $T_0 = 0.3$ and $0.01$. For $T_0 = 0.1,\ 0.03$ only frequencies that are multiples of the wind period remain. In these cases, the best stabilization is achieved. For smaller or greater values of $T_0$ "turbulent" low-frequency harmonics, which cause instability, appear again. Thus, we conclude that in the case of $\chi=17$ the periodic fluctuations of the boundary conditions with $T_0 =$ 0.03 - 0.1 give the best effect of stabilization. In the case of the Sun, it corresponds to the fluctuations with periods of 1-4 years. Analyses of solar-wind parameters show that such periodicity is observed. For example, \cite{Richardson1994} have shown that oscillations with a 1.3-year period exist.



\begin{figure*}
	\begin{minipage}[h!]{1.0\linewidth}
		\center{\includegraphics[width=1.0\linewidth]{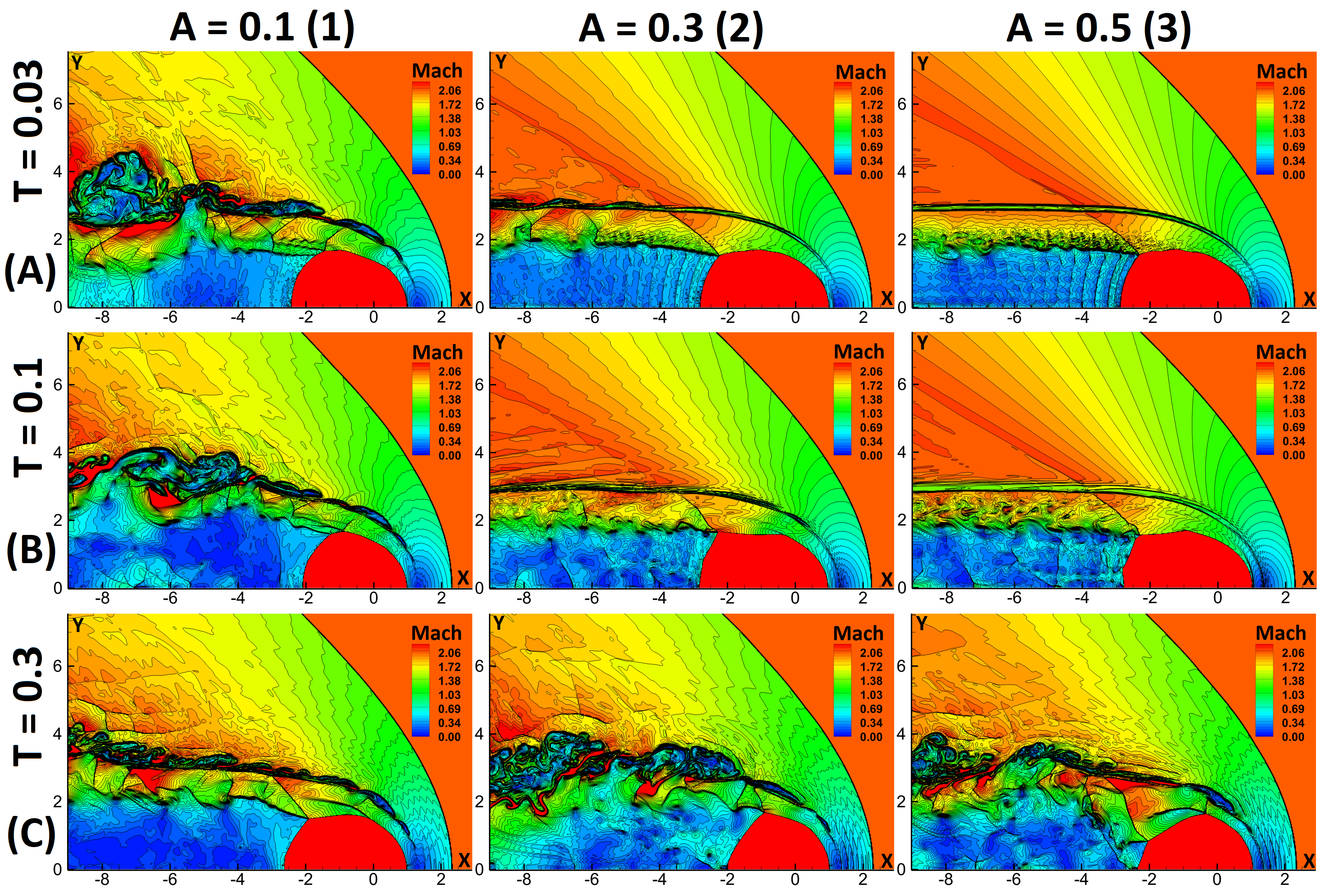} }
	\end{minipage}
	\caption{Isolines of the Mach number for different values of the stellar wind period: 0.03, 0.1, 0.3 (rows A-C) and for different values of the stellar wind amplitude: 0.1, 0.3, 0.5 (columns 1-3). $M_\infty = 2 $, $\chi = 17$, HLLC + TVD method. The grid resolution: $1792\times1536$.}
	\label{5}
\end{figure*}

Fig.~\ref{5} shows the simulation results for various amplitudes and periods of oscillations for $\chi = 17,\ M_\infty = 2$. Fig.~\ref{5} has three rows and three columns. Each column corresponds to a certain value of the solar-wind amplitude $A$. Panels in the top row (A) show the isolines of the Mach number with period $T_0 = 0.03$. We note that the cases with $A =$ 0.3 and 0.5 give the most stable flow. Such a conclusion can be drawn for $T = 0.1$ (row B). However, it is most interesting that for $T = 0.3$ the results with $A = 0.1$ show the greatest stability. It can be concluded that the best stabilization period also depends on the amplitude, while low-frequency waves can stabilize the flow at a low amplitude, and medium and high-frequency waves at a higher amplitude. We are not yet able to carry out analytical evaluations and explanations of such a result. This problem requires additional research.


In general, based on the results of two-parameter modeling, it can be concluded that high-frequency oscillations are able to stabilize tangential discontinuities in the flow, however, for too high a frequency, the discontinuities will not have time to respond to changes in parameters in the waves arriving at them, and therefore the stabilization effect disappears. Also note that in this case, the interstellar medium does not feel fluctuations in the stellar wind parameters as well.


\section{Summary and discussion}
\label{C}

The main result of the paper can be formulated as follows: Periodic variations of the stellar mass flux can reduce grows of the "turbulent" waves in the region of the stellar wind interaction with the interstellar wind.  

However, more details reported in this study can be useful during analyzes of specific astrospheres:
\begin{itemize}
\item Parametric study of the problem with stationary (not varied periodically)  stellar wind parameters has shown that the "behavior" of the astrosphere depends on the dimensionless parameter $\chi$ that is the ratio of the terminal velocity of the stellar wind to the interstellar wind speed. For small values of the parameter $\chi$ ($1 < \chi < 3$) the flow in the astrospheric boundary is laminar - no waves appeared in this case. The shape of astropause is smooth and does not vary with time.
 However, even for such small values of $\chi$ convective Kelvin-Helmholtz instability develops at the secondary tangential discontinuity in the tail region. The amplitude of fluctuations is quite small, and they convectively drifted out of the computational domain.
 
For $\chi >3$, the growth rate of K-H waves increases, and the visible fluctuations of the astropause are clearly seen in the head region. The amplitude of these fluctuations is larger for larger values of $\chi$.  For $\chi > 10$ the fluctuations of the flow in the region between the terminations shock and the heliopause significantly affect the shape of the TS making an additional breakpoint. The shock is not stationary anymore. It is position changes quasiperiodically with the main period being about 46-54 years for the heliosphere. А complex structure of refracted and reflected waves is formed in the tail region.
  
 It is important also to note that for $\chi \approx 17$ that corresponds to the heliospheric value of $\chi$ the astrospheric flow is highly fluctuating and unstable. 
 
\item The described above picture becomes much less fluctuating and disturbing when the periodic variations of the stellar wind parameters have been considered. Parametric studies by changing the period of variation $T_0$ show that the best reduction of the "turbulent" waves appears when $T_0 \approx 0.03 - 0.1$ in dimensionless units. The periodic variations with $T_0 > 0.1$ and $T_0 < 0.03$ do not damp instability waves completely. Although the reduction of their amplitudes exists for all periods.  Also, the interval of $T_0$  has been obtained for $\chi = 10-17$. For other values of $\chi$ the interval can be slightly different. In the heliospheric case, the most stabilizing effect appears for $T_0 \sim$ 1-4 year, but again, the medium-frequency waves ($4\  \mathrm{year} < T_0 < 9\ \mathrm{years}$) have a stabilizing effect in any case.  
\end{itemize}

 A possible continuation of our work is in the study of the influence of magnetic fields and neutral components on the instability of tangential discontinuities as well as the influence of the real solar cycle. 

\section*{Acknowledgements}
This research benefitted from the discussions organized by Merav Opher in the frame of SHIELD project. We also thank Dmitry Alexashov for useful discussions that stimulated this work.
The work was performed in the frame of the Russian Science Foundation grant 19-12-00383.

 \section{Data Availability}
 \label{D}
 
The videos released with this paper have been posted on Zenodo under a Creative Commons Attribution license at DOI:  \href{https://doi.org/10.5281/zenodo.7276884}{10.5281/zenodo.7276884}.

The other data underlying this article will be shared on reasonable request to the corresponding author.



\bibliographystyle{mnras}
\bibliography{mnras_template} 

\begin{thebibliography}{}
\makeatletter
\relax
\def\mn@urlcharsother{\let\do\@makeother \do\$\do\&\do\#\do\^\do\_\do\%\do\~}
\def\mn@doi{\begingroup\mn@urlcharsother \@ifnextchar [ {\mn@doi@}
  {\mn@doi@[]}}
\def\mn@doi@[#1]#2{\def\@tempa{#1}\ifx\@tempa\@empty \href
  {http://dx.doi.org/#2} {doi:#2}\else \href {http://dx.doi.org/#2} {#1}\fi
  \endgroup}
\def\mn@eprint#1#2{\mn@eprint@#1:#2::\@nil}
\def\mn@eprint@arXiv#1{\href {http://arxiv.org/abs/#1} {{\tt arXiv:#1}}}
\def\mn@eprint@dblp#1{\href {http://dblp.uni-trier.de/rec/bibtex/#1.xml}
  {dblp:#1}}
\def\mn@eprint@#1:#2:#3:#4\@nil{\def\@tempa {#1}\def\@tempb {#2}\def\@tempc
  {#3}\ifx \@tempc \@empty \let \@tempc \@tempb \let \@tempb \@tempa \fi \ifx
  \@tempb \@empty \def\@tempb {arXiv}\fi \@ifundefined
  {mn@eprint@\@tempb}{\@tempb:\@tempc}{\expandafter \expandafter \csname
  mn@eprint@\@tempb\endcsname \expandafter{\@tempc}}}

\bibitem[\protect\citeauthoryear{Baranov, Krasnobaev  \& Kulikovskii}{Baranov
  et~al.}{1970}]{baranov70}
Baranov V.~B.,  Krasnobaev K.~V.,   Kulikovskii A.~G.,  1970, Ser.~Mat.~Fiz.,
  41, 194

\bibitem[\protect\citeauthoryear{{Baranov}, {Fahr}  \& {Ruderman}}{{Baranov}
  et~al.}{1992}]{Baranov1992}
{Baranov} V.~B.,  {Fahr} H.~J.,   {Ruderman} M.~S.,  1992, \aap, \href
  {https://ui.adsabs.harvard.edu/abs/1992A&A...261..341B} {261, 341}

\bibitem[\protect\citeauthoryear{{Chalov}}{{Chalov}}{1996}]{Chalov1996}
{Chalov} S.~V.,  1996, \aap, \href
  {https://ui.adsabs.harvard.edu/abs/1996A&A...308..995C} {308, 995}

\bibitem[\protect\citeauthoryear{Cox et~al.,}{Cox et~al.}{2012}]{Cox_2012}
Cox N. L.~J.,  et~al., 2012, \mn@doi [A\&A] {10.1051/0004-6361/201117910}, 537

\bibitem[\protect\citeauthoryear{Decin et~al.,}{Decin et~al.}{2012}]{Decin}
Decin L.,  et~al., 2012, \mn@doi [A\&A] {10.1051/0004-6361/201219792}, 548

\bibitem[\protect\citeauthoryear{{Dgani}, {van Buren}  \&
  {Noriega-Crespo}}{{Dgani} et~al.}{1996}]{Dgani}
{Dgani} R.,  {van Buren} D.,   {Noriega-Crespo} A.,  1996, \mn@doi [\apj]
  {10.1086/177114}, \href
  {https://ui.adsabs.harvard.edu/abs/1996ApJ...461..927D} {461, 927}

\bibitem[\protect\citeauthoryear{Gurski}{Gurski}{2004}]{Gurski}
Gurski K.,  2004, SJOCE, 2165, 25

\bibitem[\protect\citeauthoryear{Helmholtz}{Helmholtz}{1868}]{Helmholtz}
Helmholtz H.,  1868, Monthly reports of the Royal Prussian Academy of Sciences
  in Berlin, 23, 215–228

\bibitem[\protect\citeauthoryear{{Herbst} et~al.,}{{Herbst}
  et~al.}{2022}]{Herbst}
{Herbst} K.,  et~al., 2022, \mn@doi [\ssr] {10.1007/s11214-022-00894-3}, \href
  {https://ui.adsabs.harvard.edu/abs/2022SSRv..218...29H} {218, 29}

\bibitem[\protect\citeauthoryear{Izmodenov, Malama  \& Ruderman}{Izmodenov
  et~al.}{2005}]{Izmodenov2005}
Izmodenov V.,  Malama Y.,   Ruderman M.~S.,  2005, \mn@doi [A\&A]
  {10.1051/0004-6361:20041348}, 429, 1069–1080

\bibitem[\protect\citeauthoryear{Kapitza}{Kapitza}{1951a}]{Kapitza1}
Kapitza P.~L.,  1951a, Soviet Phys., 21, 588–597

\bibitem[\protect\citeauthoryear{Kapitza}{Kapitza}{1951b}]{Kapitza2}
Kapitza P.~L.,  1951b, \mn@doi [Usp. Fiz. Nauk.]
  {10.3367/UFNr.0044.195105b.0007}, 44, 7–15

\bibitem[\protect\citeauthoryear{Kelvin}{Kelvin}{1871}]{Kelvin}
Kelvin 1871, Philosophical Magazine, 42, 362–377

\bibitem[\protect\citeauthoryear{Kobulnicky et~al.,}{Kobulnicky
  et~al.}{2016}]{Kobul_2016}
Kobulnicky H.~A.,  et~al., 2016, \mn@doi [ApJS] {10.3847/0067-0049/227/2/18},
  227

\bibitem[\protect\citeauthoryear{{Kobulnicky}, {Schurhammer}, {Baldwin},
  {Chick}, {Dixon}, {Lee}  \& {Povich}}{{Kobulnicky} et~al.}{2017}]{Kobul_2017}
{Kobulnicky} H.~A.,  {Schurhammer} D.~P.,  {Baldwin} D.~J.,  {Chick} W.~T.,
  {Dixon} D.~M.,  {Lee} D.,   {Povich} M.~S.,  2017, \mn@doi [\aj]
  {10.3847/1538-3881/aa90ba}, \href
  {https://ui.adsabs.harvard.edu/abs/2017AJ....154..201K} {154, 201}

\bibitem[\protect\citeauthoryear{Korolkov, Izmodenov  \& Alexashov}{Korolkov
  et~al.}{2020}]{Korolkov_2020}
Korolkov S.~D.,  Izmodenov V.~V.,   Alexashov D.~B.,  2020, \mn@doi
  [J.Phys.:Conf.Ser] {10.1088/1742-6596/1640/1/012012}

\bibitem[\protect\citeauthoryear{Kulikovski \& Shikina}{Kulikovski \&
  Shikina}{1977}]{Kulikovski1977}
Kulikovski A.~G.,  Shikina I.~C.,  1977, 12, 679–682

\bibitem[\protect\citeauthoryear{Landau \& Lifshitz}{Landau \&
  Lifshitz}{1960}]{Landau1}
Landau L.,  Lifshitz E.,  1960, Mechanics.
No.~1, Pergamon Press, \url {https://www.amazon.com/dp/B0006AWV88}

\bibitem[\protect\citeauthoryear{Landau \& Lifshitz}{Landau \&
  Lifshitz}{2013}]{Landau6}
Landau L.,  Lifshitz E.,  2013, Fluid Mechanics: Volume 6.
No.~6, Elsevier Science, \url {https://books.google.ru/books?id=CeBbAwAAQBAJ}

\bibitem[\protect\citeauthoryear{{Meyer}, {Gvaramadze}, {Langer}, {Mackey},
  {Boumis}  \& {Mohamed}}{{Meyer} et~al.}{2014a}]{Meyer2014}
{Meyer} D.~M.~A.,  {Gvaramadze} V.~V.,  {Langer} N.,  {Mackey} J.,  {Boumis}
  P.,   {Mohamed} S.,  2014a, \mn@doi [\mnras] {10.1093/mnrasl/slt176}, \href
  {https://ui.adsabs.harvard.edu/abs/2014MNRAS.439L..41M} {439, L41}

\bibitem[\protect\citeauthoryear{Meyer, Mackey, Langer, Gvaramadze, Mignone,
  Izzard  \& Kaper}{Meyer et~al.}{2014b}]{Meyer}
Meyer D. M.-A.,  Mackey J.,  Langer N.,  Gvaramadze V.~V.,  Mignone A.,  Izzard
  R.~G.,   Kaper L.,  2014b, \mn@doi [MNRAS] {doi:10.1093/mnras/stu1629}, 444,
  2754

\bibitem[\protect\citeauthoryear{{Meyer}, {Mignone}, {Petrov}, {Scherer},
  {Vel{\'a}zquez}  \& {Boumis}}{{Meyer} et~al.}{2021}]{Meyer2021}
{Meyer} D.~M.~A.,  {Mignone} A.,  {Petrov} M.,  {Scherer} K.,  {Vel{\'a}zquez}
  P.~F.,   {Boumis} P.,  2021, \mn@doi [\mnras] {10.1093/mnras/stab2026}, \href
  {https://ui.adsabs.harvard.edu/abs/2021MNRAS.506.5170M} {506, 5170}

\bibitem[\protect\citeauthoryear{Miyoshi \& Kusano}{Miyoshi \&
  Kusano}{2005}]{Miyoshi}
Miyoshi T.,  Kusano K.,  2005, \mn@doi [J.Comp.Phys]
  {10.1016/j.jcp.2005.02.017}, 208, 315

\bibitem[\protect\citeauthoryear{{Noriega-Crespo}, {van Buren}, {Cao}  \&
  {Dgani}}{{Noriega-Crespo} et~al.}{1997}]{Noriega}
{Noriega-Crespo} A.,  {van Buren} D.,  {Cao} Y.,   {Dgani} R.,  1997, \mn@doi
  [\aj] {10.1086/118517}, \href
  {https://ui.adsabs.harvard.edu/abs/1997AJ....114..837N} {114, 837}

\bibitem[\protect\citeauthoryear{{Richardson}, {Paularena}, {Belcher}  \&
  {Lazarus}}{{Richardson} et~al.}{1994}]{Richardson1994}
{Richardson} J.~D.,  {Paularena} K.~I.,  {Belcher} J.~W.,   {Lazarus} A.~J.,
  1994, \mn@doi [\grl] {10.1029/94GL01076}, \href
  {https://ui.adsabs.harvard.edu/abs/1994GeoRL..21.1559R} {21, 1559}

\bibitem[\protect\citeauthoryear{{Ruderman} \& {Fahr}}{{Ruderman} \&
  {Fahr}}{1993}]{Ruderman1993}
{Ruderman} M.~S.,  {Fahr} H.~J.,  1993, \aap, \href
  {https://ui.adsabs.harvard.edu/abs/1993A&A...275..635R} {275, 635}

\bibitem[\protect\citeauthoryear{Ruderman, Brevdo  \& Erdelyi}{Ruderman
  et~al.}{2004}]{Ruderman}
Ruderman M.~S.,  Brevdo L.,   Erdelyi R.,  2004

\bibitem[\protect\citeauthoryear{{\noopsort{V}van Buren} \&
  {McCray}}{{\noopsort{V}van Buren} \& {McCray}}{1988}]{Buren1988}
{\noopsort{V}van Buren} D.,  {McCray} R.,  1988, \mn@doi [\apjl]
  {10.1086/185184}, \href
  {https://ui.adsabs.harvard.edu/abs/1988ApJ...329L..93V} {329, L93}

\bibitem[\protect\citeauthoryear{{\noopsort{V}van Buren}, {Noriega-Crespo}  \&
  {Dgani}}{{\noopsort{V}van Buren} et~al.}{1995}]{Buren1995}
{\noopsort{V}van Buren} D.,  {Noriega-Crespo} A.,   {Dgani} R.,  1995, \mn@doi
  [\aj] {10.1086/117739}, \href
  {https://ui.adsabs.harvard.edu/abs/1995AJ....110.2914V} {110, 2914}

\bibitem[\protect\citeauthoryear{{Wallis} \& {Dryer}}{{Wallis} \&
  {Dryer}}{1976}]{Wallis1976}
{Wallis} M.~K.,  {Dryer} M.,  1976, \mn@doi [\apj] {10.1086/154345}, \href
  {https://ui.adsabs.harvard.edu/abs/1976ApJ...205..895W} {205, 895}

\bibitem[\protect\citeauthoryear{{Wang} \& {Belcher}}{{Wang} \&
  {Belcher}}{1998}]{Wang1998}
{Wang} C.,  {Belcher} J.~W.,  1998, \mn@doi [\jgr] {10.1029/97JA02773}, \href
  {https://ui.adsabs.harvard.edu/abs/1998JGR...103..247W} {103, 247}

\bibitem[\protect\citeauthoryear{{Wilkin}}{{Wilkin}}{1996}]{Wilkin}
{Wilkin} F.~P.,  1996, \mn@doi [\apjl] {10.1086/309939}, \href
  {https://ui.adsabs.harvard.edu/abs/1996ApJ...459L..31W} {459, L31}

\makeatother
\end{thebibliography}




\bsp	
\label{lastpage}
\end{document}